# CyberOps: Situational Awareness in Cybersecurity Operations

**Cyril Onwubiko**

*Enterprise Security Architecture, Pearson Plc*
*Artificial Intelligence, Blockchain & Cyber Security, Research Series, London, UK*

## ABSTRACT

Cybersecurity operations (CyberOps) is the use and application of cybersecurity capabilities to a domain, department, organisation or nation. It is fundamentally to protect digital investments, contribute to national economic wellbeing by providing a safe, secure and conducive environment to conduct business and to protect a nation's critical national infrastructures and citizens welfare. In this paper, we investigate operational factors that influence situational awareness of CyberOps, specifically, the features that deals with understanding and comprehension of operational and human factors aspects and that helps with insights on human operator decision making (e.g., cognition, teamwork, knowledge, skills and abilities). The operational factors discussed in this paper range from tools, techniques, integration, architecture to automation, cognition, people, policy, process and procedures.

## 1    INTRODUCTION

Like engineering, computer science or chemistry, cybersecurity is now considered a mainstream discipline. Its importance has never been felt more than the past 12 months when the world has been forced to adapt to a 'new normal' as a result of the COVID-19 pandemic. This 'new normal' has forced people to remote-everything, from remote working, remote tutoring, remote schooling, to remote socials, e.g., 'Zoom lunch', 'Zoom Christmas'.

Despite the many benefits derived from remote and online activities, but the 'new normal' is not without its challenges and continues to be tested on a daily basis. Whether through the increased attacks on university learning platforms and theft of students' data (BBC, 2021a), attacks on online banking services, to nation-state malware embedded in school pupils' laptop provided by government (BBC, 2021b).

Cyber-attacks have become sophisticated and increasing in number. The sophistication is as a result of advances in technology, automation and emerging techniques, for example, advances in machine learning and artificial intelligence have been leveraged to conduct deepfake (exact copy look-alike) of genuine emails, transactions, audio and video transcripts. We do not suggest that the several folds increase in cyber-attacks in the last 12 months of the COVID-19 pandemic is as a result of sophistication in technology, and neither are we suggesting that all the recent cyber-attacks have used sophisticated techniques. Indeed, some recent attacks have used sophisticated techniques to evade detection, but the fact remains, majority used the same 'old' techniques that have been in the wild for several years. Further, these attacks do not target one type of organisation, a particular vertical, country or region. We have seen accounts of cyber-attacks to government and military, health, banking, education sectors and individual and home users.

One thing is evident, we need appropriate cybersecurity capabilities to secure individual, organisation and national digital investments, and one way of doing this is through cybersecurity operations (CyberOps). Hence, the motivation for this paper is to investigate CyberOps. Specifically, best practices and the factors, which influence their situational awareness.

The key contributions of this manuscript can be summarised as:

1) Introduce CyberOps in the wider context and discuss its relationship to constituent components/capabilities.
2) Explain Zero Trust with respect to entity interaction diagram (see Fig. 2) and coarse-level principles.
3) Provide insights to Cyber Situational Awareness (CyberSA).
4) Introduce and explain our proposed BOTH (Business, Operation, Technology and Human) factors.
5) Investigate operational factors influencing situational awareness of the analysts in CyberOps.

This paper is organised as follows. Section 2 provides insights into cybersecurity operations. It discusses CyberOps, the difference between CyberOps and CSOC (a.k.a. SOC); introduces Zero Trust Architecture and outlines related works. We then explain situational awareness in CyberOps in Section 3. This is followed by an in-depth explanation of factors affecting cyber situational awareness in CyberOps. Each of the factors are explained, followed with an example or a discussion. In Section 5 we discuss future work, while Section 6 concludes the paper.

## 2 UNDERSTANDING CYBEROPS

### 2.1 CyberOps

CyberOps (a.k.a. Cyber Security Operations, Cyber Operations, Security Operations, SecOps) is simply, the use or application of Cyber capabilities in a particular domain.

For example, the application of Cyber capabilities in the Military domain, is called, Military Cyber Operations (see Schulze M., 2020); the application of Cyber capabilities in Operational Technology is referred to as (Cyber OT), e.g., Industrial Control Systems, Supervisory Control and Data Acquisition (SCADA) and Distributed Control Systems (see (NCSC, 2021); (ENISA, 2016)); the application of Cyber capabilities for intelligence purpose, is regarded as Cyber intelligence operation (Intel Op), while the use of Cyber capabilities for threat intelligence purposes is regarded as Cyber threat intel (CTI). The application of Cyber capabilities in information technology (IT) and information communications technology (ICT), i.e., cyber in ICT is generally regarded as Cyber Operations.

CyberOps, the use of Cyber capabilities in ICT, is discussed in relation to the functions, operations (e.g., activities, tasks) and responsibilities performed to *administer*, *operate, monitor* and *support* cybersecurity services, systems, applications, platforms and infrastructures to fulfil business goals.

It is important to note that CyberOps is very broad, encompassing four key responsibilities: 1) Administration, 2) Execution, 3) Monitoring and 4) Support. These responsibilities can be accomplished by a single team or multiple teams. The teams can either be centralized or distributed and diverse. The choice as to whether the organisation operates a centralized or distributed team structure is dependent on many factors, such as, size, structure, business model, investment, and business operating model of the organisation, institution (e.g., Government or Agency) or/and establishment.

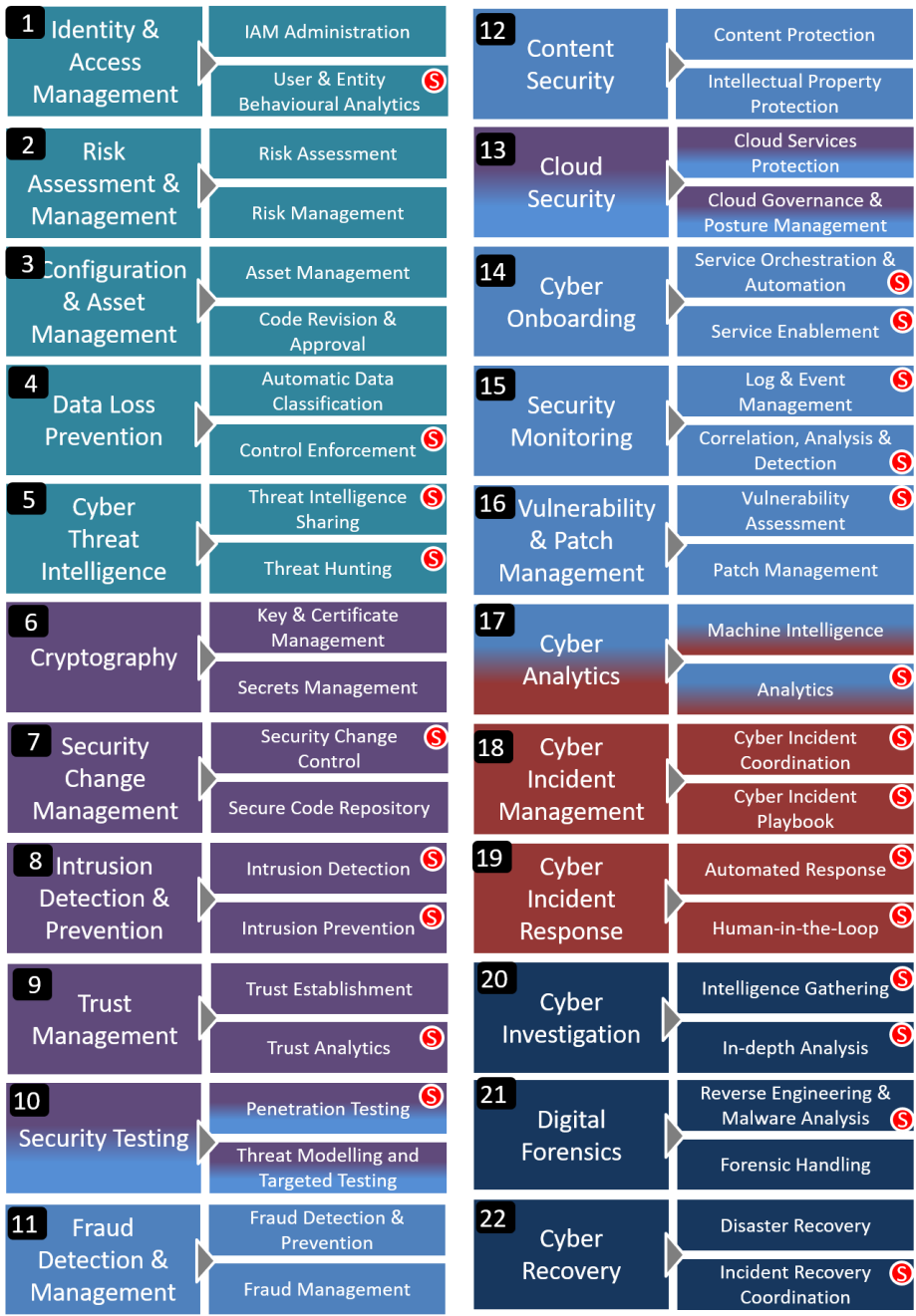

FIGURE 1: CYBEROPS FUNCTIONAL REPRESENTATION

Since CyberOps is broad and diverse, it is challenging to comprehensively describe or define it. We describe it to cover not just the administration and execution, but also the monitoring and support of cybersecurity systems, applications and services. That is, the administration, execution, monitoring and supporting of an organisation's *protect surface* – the *data, assets, applications* and *services* (DAAS) – to ensure errors, incidents, faults and failures are identified, detected and remedied. We use the word '*Monitoring*' in this text to comprise *detection* through to *response*, *remediation*, and *recovery*.

These tasks can be performed by one team or several teams within the organisation, and in some cases, the responsibility could be outsourced to a security and/or service provider organisation to perform. The decision as to whether one team or several teams perform the cybersecurity operations functions, and whether it is outsourced to a service provider organisation is dependent on several factors. For example, the size of the organisation, their operating model, vision and cybersecurity strategy.

In Fig .1 we represent key constituent capabilities of a typical CyberOps within an organisation. These capabilities are broken down at a very high level into subcomponents, at most, two subcomponents for the sake of brevity. These capabilities are mapped to the National Institute of Standards and Technology (NIST) Cybersecurity Framework (NIST CSF, 2021). We recognise that a capability may address multiple issues and may be aligned to a couple of domains (that is in respect to the NIST CSF, of identify, protect, detect, respond and recover), hence where we believe this to be the case, we colour-coded the subcomponents of the capability in overlapping colours. For example, capabilities #10 (Security Testing), #13 (Cloud Security) and #17 (Cyber Analytics) are divided primarily into two. Note: *These capabilities can be split into multiple granular levels, but to aid understanding, we partitioned each capability to no more than two.*

The alignment and structuring of these capabilities may differ from organisation to organisation. For example:
a) One organisation may combine Security Testing (Cap. #10), with Vulnerability and Patch Management (Cap. #16) under one function or accountability, and many of the capabilities may be structured differently across organisation for many reasons, such as cost, size of the organisation, cyber maturity of the organisation, scale of transactions or business.
b) We do not expect every organisation to rollout (or implement) all the capabilities at once, or even at all for obvious reasons, for example, an organisation may be constrained by resources (e.g., skilled manpower), cost, and business necessities.

c) Some of the capabilities should be combined in some organisations due to the reasons provided in (a) above.

Each capability represented in Fig. 1 encompasses four aspects to realise it, namely: 1) Architecture/Engineering, 2) Systems Administration, 3) Systems Operation and 4) Systems Support. *Architecture and engineering* deals with the design, the building blocks and the blueprint that stems from business requirements and needs for the capability. From the design the capability is built. *Administration* deals with configuration, setting up and fine tuning the capability such that it aligns to the architecture or engineering design and to realise both the functional and nonfunctional requirements, and does include adding new instances and users of such capability. *Operation* deals with operating the capability, monitoring the capability to ensure it works according to how it is built, and that faults, errors, failures and incidents are detected, remediated and services restored to normal, and finally, *Support* deals breaks and fixes, ensuring the faults and failures detected on the capability are remedied and that services are restored back to normal operating state.

The effectiveness and efficiency of CyberOps should be measured in relation to a number of factors such as tools, techniques, process maturity, service automation and orchestration, knowledge, skills and abilities (KSA) of the analysts, operators and administrators of the service.

## 2.1    Cyber Security Operations Centre (CSOC or SOC)

According to Carson Zimmerman (Zimmerman, C. 2014), a SOC is a team of people comprising security analysts who perform detection, analysis, response, reporting and prevention of cybersecurity incidents. The functions that the SOC performs can also be performed by teams who may be known by other names, such as the Computer Emergency Response Team (CERT), Computer Security Incident Response Team (CSIRT), Computer Incident Response Team (CIRT) etc. these terms may be used interchangeably in some publications.

We describe SOC as a *horizontal business function* of an organisation, institution, industry comprising people, technology and process. They are responsible for continuous security monitoring, cyber incident response, cyber security incident management, detection, monitoring, log and event management. We describe it as a horizontal business function because the effectiveness of a SOC to the organisation is best realised when the responsibilities of the SOC is for the entire organisation, as opposed to multiple, tactical, isolated, standalone and fragmented SOCs that lack situational awareness of the risks the organisation bears as a whole.

CSOC, SOC, fusion centres and CIRTS or CERTs are a subset of CyberOps. It is imperative to understand that CyberOps encompass all the disparate, and sometimes related cyber capabilities performed by the different teams, ranging from identification, protection, detection, monitoring through to respond and recovery from cyber-attacks.

Another distinction worth noting is that cybersecurity operations is not the same as cyber security operations center (CSOC). CSOC or Security Operations Centre (SOC) is a subset of CyberOps. CyberOps is a much broader capability (as seen in Fig. 1) than SOC, and they perform much more responsibilities than a SOC or a CSOC normally does.

As shown in Fig. 1, the capabilities denoted with  can be said to belong to a typical SOC, while the other capabilities are performed by the wider security organisation and not by the SOC. In fact, in some organisations, one could argue that some of the capabilities denoted with  could be performed by other security organisation. For example, some organisations have Cyber Threat Intelligence and Cyber & Data Analytics dedicated Team / Group or Departments and as such these functions are performed by these Teams / Groups or Departments, but not necessarily by the SOC.

## 2.2 Zero Trust

We recognise that an effective SOC should have an enriched monitoring platform, capable of providing the requisite situational awareness of the monitored *protect surface*, comprising Data, Assets, Applications and Services (DAAS). It is pertinent to note that *protect surface*, a Zero Trust concept, formulated by John Kindervag (Kindervag J., 2020) comprises an organisation critical data (e.g., personally identified information (PII)), assets, applications and services they wish to prioritise for protection and security monitoring. Since not all DAAS of an organisation warrants the same level of protection, and neither could protection be provided to every service at once, it is important to prioritise which 'surfaces are protected'. In other words, which critical services of the organisation must be protected, especially through an informed risk proportionality assessment approach? Understanding the protect surface is the key starting point to providing efficient protection.

Zero Trust is a descriptive concept, a set of security and architecture principles on how best to protect critical resources of an organisation (e.g., DAAS) and how access to these can be safeguarded. It is underpinned on seven (7) high-level principles, which are discussed in Table 1.

TABLE 1: ZERO TRUST PRINCIPLES

| | Principles | Description |
|---|---|---|
| 1 | *Non-Presumptive Access* | Access and rights are neither static nor assumed. Instead access, privileges and rights are determined based on the level of risk and trust assessed of the entity and identity requesting the access. |
| 2 | *Zero Trust Architecture (ZTA)* | Design the network architecture to align to business requirements, ensuring that access to business-critical services (DAAS) is less dependent on, and transparent to, the network infrastructure and security is a function of the identity rather than 'hard border' or supposedly network boundary. |
| 3 | *Conditional Access (a.k.a. adaptive risk and trust-based) assessment* | All interactions from workforce (e.g., user, persona or identity and endpoint) to workload and workplace (e.g., Data, Assets, Applications, and Services) are risk and trust assessed throughout the duration of the interaction. See Section 2.3, conditions 2a-2f. Adaptive risk and trust assessment is the underlining construct to achieving a Zero Trust assessment of Gartner's CARTA – continuous adaptive risk and trust assessment (Gartner, 2021). |
| 4 | *Context-aware security* | Access to targeted resources e.g., DAAS – Data, Asset, Application and Service from all identities are controlled based on context. For example, the '*what*', from '*where*' and '*why*' and for '*how*' long. |
| 5 | *Secure Access* | The focus for secure access is underpinned on the principle of *least-privilege access* combined with *multi-factor authentication* (MFA) and *certificate-based endpoint protection*. Ensure overprivileged accounts are rationalized, MFA is rolled out, and managed endpoints are cert-based. |

| 6 | *Identity is the 'new' Perimeter (Perimeterless)* | Assume no "North-South" perimeters anymore. Perimeters will gradually extend to the very identities (e.g., users, devices, applications and workload) they protect, e.g., edge computing, and edge protection. This is the same principle as 'Application and identity-centric protection over infrastructure and network-based protection'. |
| 7 | *Secure the Protect Surface* | Since not all services of an organisation requires the same level of protection, and in reality, not every single DAAS of an organisation may be adequately protected, therefore, identifying and prioritising the protect surface is essential. Knowing your critical data, assets, applications and services, how they can be accessed, by whom are extremely valuable. Since access to DAAS can be initiated from outside and within, and there is no differentiation. |

## 2.2.1 Understanding Zero Trust Architecture

An Internet Protocol (IP)-based communication starts from a source, usually an identity using an entity or endpoint through a network (IP-based network) to a target, usually a business services, such as a web server, database or an application. The network encapsulates many resources such as switch, router, firewall, intrusion detection system etc. These resources perform a series of functions, ranging from routing, filtering, inspection to forwarding. This is how the source gets to the target. The source and the service it wants to access (a.k.a. target) can be in the same organisation, in which case, it is internal to internal communication. On the other hand, the source can be external while the target is internal or vice versa, in which case is external to internal communication.

Fig 2 is our representation of a conceptual Zero Trust Architecture. It shows how a source (**A**) uses an entity or endpoint (**B**) through an abstracted network (**C**) accesses a business service (**D**). We use the notation of an 'abstracted network' to imply that the network is simply abridged, meaning a paradigm shift from securing the perimeter to securing the identity (e.g., user or persona). We argue that 'identity'-based protection offers robust security over perimeter-defence since better understanding of the risks presented by the identity and the endpoint is far more prudent than 'perimeter' focused-defence.

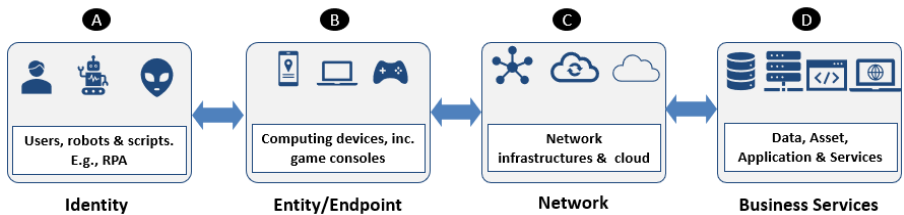

FIGURE 2: ZERO TRUST REPRESENTATION

As shown in Fig. 2, imagine *identities* (e.g., user persona, robots, scripts) as shown in (**A**) through *entities* (e.g., Tablets, Laptops etc.) as shown in (**B**) requesting access to *Business Services* as shown in (**D**) through *Network Services* as shown in (**C**).

Conditional access is predicated on the continuous assessment of risks and trust of identity and entity (see Fig. 2) requesting access to critical services may poss. These conditions are essential for achieving conditional access:

a)  Could the identity and endpoint be used to compromise the service?
b)  Could the entity (e.g., health hygiene) be in such a condition e.g., malware infested, such that through it the service it may be requested may be infested with malware?
c)  Could the entity pose material risks, geographical risks (e.g., geolocation) or regulatory risks (e.g., access from certain geographies that are under a sanction or prohibited from accessing the service)?
d)  Could there be underlying risk (e.g., backdoor access) that this entity may introduce?
e)  Could the entity or identity be used for financial extortion?
f)  Could the access originate from identity and/or entity from a geography (geo-location) that is prohibited or untrusted (e.g., risky)
g)  Is the behaviour of the identity and entity symptomatic of an attacker (e.g., odd, unusual, or suspicious behaviour), e.g., access at odd times, random access to several services that the identity does not have access to, and/or should not be accessing e.g., hidden files, protected documents etc.

Zero Trust is principally focused on non-presumptive access, that is, not granting any identity or entity presumptive access, rights or privileges to critical services, instead access to critical services should be continuously assessed based on trust and risk (a.k.a. conditional access) of the requesting identity. It could also be likened to preserving the least privilege principle of security protection, where access given to identities requesting services must be the least minimum required to complete the work rather than granting overly permissive access to an identity.

## 2.3 Related Works

In the literature, attempts to describe and discuss Cyber Operations exist.

The Cyber Security Body of Knowledge (CyBOK), an initiative to inform and align cyber security to thematic domains for education and professional training (CyBOK, 2019) offers insights. Through the CyBOK initiative, Herve Debar (Debar H., 2019) describes Security Operations and Incident Management (SOIM) as the application and automation of the Monitor, Analyse, Plan, Execute-Knowledge (MAPE-K) autonomic computing loop to cybersecurity. This description while combining incident management focused primarily on the constituent aspects of the monitoring, detection, response and remediation aspects of cyber security operations.

Schulze M. (2020) discusses Cyber Operations in relation to the military domain, where cyber capabilities are discussed in respect to offensive and defensive operations, although the motivation for his contribution focuses on the use of Cyber Operations in War, that is, the active use of cyber capabilities in war time. It follows a set of coordinated actions with very precisely defined military purpose in cyberspace. The use of cyber capabilities in an active war is not only to prove superiority over your adversary, but also, to deny your adversary the ability to respond, influence or succeed in that war.

The use of Cyber capabilities in Maritime (i.e., the maritime section) is an emerging area of recent contributions. For example, Kimberly Tam and Kevin D Jones (Tam K. & Jones K. D., 2019) investigate the pertinent need for realtime and adaptive cyber risk assessment and situational awareness in maritime. Their work investigates the need for cyber operations in maritime for timely, realtime and adaptive risk assessment and situational awareness in managing cyber-physical risks to the maritime sector. Modern ships are built these days with extensive reliance on advance operational technology (OT), e.g., onboard circuitry, chips and embedded systems, and these have huge reliance on information technology for receiving information and signals, radar communications, e.g., from the global navigation satellite systems (GNSS), and various other situational awareness cues and prompts that are provided to operators onboard. Understanding the conjecture and relationship between OT and IT in maritime and the bridge between OT and IT is essential to effective cyber operations in maritime (Man, Lundh & MacKinnon, 2018).

Aviation Cyber Security focuses on the use of cybersecurity capabilities to ensure safe, secure and resilient operation in aeronautics (UK Aviation Cyber Security Strategy, 2018). With the increasing number of cyber threats to modern aircrafts and the proliferation of vulnerabilities in IT systems, CyberOps and situational awareness in this space is extremely important. According to the Military Aviation Authority (MAA), modern military aircrafts and their supporting ground systems (e.g., Air Traffic Controls, Operators etc.) are now hugely reliant on computer systems for safe and efficient operations making them susceptible to cyber-attacks (MAA, 2020). It is recognised that CyberOps in aviation is extremely essential to counter cyber-attacks that poses significant threats the safe and operation of modern aircrafts and its supporting ground systems.

Our work provides a holistic understanding of CyberOps drawn from the contributions of previous body of knowledge cited in this article.

## 3    CYBER SITUATIONAL AWARENESS

Since the seminal work of Endsley (Endsley, 1995) situational awareness has been applied to a number of areas such as safety, security and transportation. Situational awareness has been applied in cyber security, vehicular networks, aviation, social media analytics and conversational agents (Onwubiko C., 2009, Eiza M. H., 2017, & McDermott, C. D., Jeannelle, B., and Isaacs, J. P., 2019).

*Cyber Situational Awareness (Cyber SA)* has been defined in many ways in relation to cyber security, cyber defense, and cyber operations in general (see, McGuiness B., Foy J. L., 2000, Cumiford D. L., 2006 and Tadda G. P., and Salerno J. S., 2010). We adopt the definition of Cyber SA provided by Onwubiko and Owens (Onwubiko C. & Owens T., 2012), which states that "situational awareness is the ability of being aware of circumstances that exist around us, especially those that are particularly relevant to us and which we are interested about. It encompasses the prediction of future states or impending states as a result of the knowledge (which could include new information) and understanding of current states".

Situational awareness is ideal for understanding operational and human factors aspects and helps with insights on human operator decision making (e.g., cognition, teamwork, knowledge, skills and abilities). We see this to be pertinent in this paper, especially in gaining 'understanding' of the relationships through humans-in-the-loop. Humans-in-the-loop are Cyber Security Operators (e.g., Analysts, Administrators, Scientists, Engineers etc.) who take inputs from technologies (e.g., security enforcing functions devices,

SIEM etc.) to aid decision making. They understand and have experience of cyber security incident management and assessment, major incident management, detection, cyber incident response and recovery. These humans-in-the-loop leverage technology, automation and integration combined with their experience, skills and knowledge provide cyber foresight. Further and most importantly, the interdependence and inter-dimensionality of the multiple domains e.g., physical, cultural, economic, social, political and cyber that must be considered in order that enhanced situational awareness across the domains can be achieved.

Cyber defence tools are not a 'silver bullet', and do not solve all the cyber security problems themselves. For example, cyber defence tools such as firewalls or intrusion detection systems are unable to solve cyber security procedural or human factors problems. They are as efficient as the people who use them to monitor business services, follow up on incidents and conduct incident triage. The tools may offer cues and prompts which the human operators, such as cyber security operations centre (CSOC) administrators and analysts should investigate. Often these cues are *symptomatic* - an expression of a likelihood of something, rather than an explicit indication, therefore, human expertise and experience are very much required. The cues which are provided by the monitoring systems may be in the form of alerts, alarms, events etc. These intelligences will need to be analysed and decision on possible cause of action (CoA) will be down to humans to make.

*Cyber situations* include cyber threats, cyber security attacks, cyber risks and cyber issues, such as cyber vulnerabilities, exploits, security breaches and cybercrime.

## 4    OPERATIONAL FACTORS INFLUENCING CYBER SITUATIONAL AWARENESS IN CYBER OPERATIONS

In this section, we investigate factors that influence operators' situational awareness in CyberOps, specifically operational factors. The term 'operator' is used here as a generalistic term to describe administrators, analysts, scientists, engineers and architects whose roles in cybersecurity involves administering, operating, monitoring and supporting cybersecurity of an organisation, as discussed in Section 2 of this paper, and functionally represented in Fig 1.

*Operational factors* have been described and discussed in many ways. In logistics, operational factors are defined as factors that are used to evaluate alternatives capabilities for meeting the external requirements of outbound logistics services (Çebi, F., Otay, İ., & Çelebi, D. 2014); and according to Reverso Dictionary (Reverso, 2021), operational factors relate to the working of a system, device or a plan. In this paper, we adopt the Reverso dictionary definition when investigating operational factors, namely those factors relating to, and that influence the working of cybersecurity systems, devices and/or their plans, policies, processes and procedures.

Further, we investigate how the operational factors influence operators' situational awareness in CyberOps. How do these operational factors help the operators:

- Identify and detect threats and cyber-attacks?
- Respond and manage the affected systems and services when they are breached or compromised?
- Recover and remedy the systems and services when they fail?

In general, we propose four overarching group of factors that we considered to influence cyber situational awareness. These are **business**, **operational, technology** and **human factors (a.k.a. BOTH factors)**. They are derived using attribute listing, matrix and morphological analysis methodology.

Attribute listing and morphological analysis is one of the methods applied in this research to enumerate all possible attributes and features stemming from each domain we considered. For example, to understand operational factors that influence cyber situational awareness in cybersecurity operations we listed all possible operational attributes and feature sets of operational tasks involved in cybersecurity domain. By further deduplicating and pruning identical and redundant attributes and features we then obtained a subset of the initial list. This iteration happened several times until a baseline is obtained. The baseline is as shown in Fig. 3.

*Security, privacy* and *information assurance* requirements neither appeared in any specific quadrant nor included in any of the factors. We argue that information security, privacy and information assurance are intrinsic features that must be considered in their own rights for all the four principal factors. Therefore, these requirements are embedded across the four principal factors (as shown in Fig. 3), e.g., Business Factors, Operational Factors, Technological Factors and Human Factors.

For example, operations factors must consider information security, privacy and information assurance requirements with respect to the operations that the tools undertake and the processes must be such that they align to security standards and privacy regulations and directives. Similarly, business, technology and human factors must all abide by the same principles.

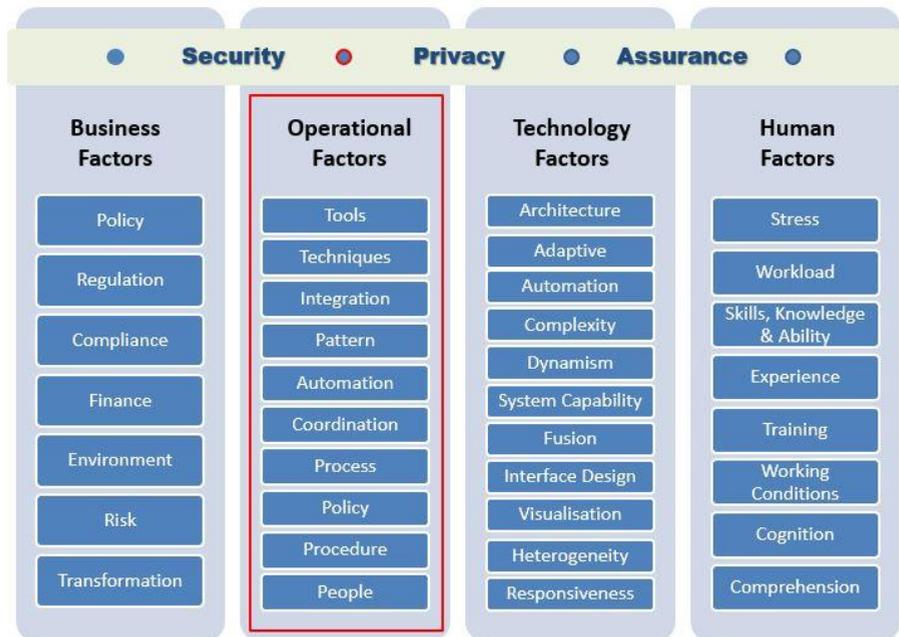

FIGURE 3: BOTH – BUSINESS, OPERATIONAL, TECHNOLOGY & HUMAN FACTORS

We identified ten (10) pertinent operational factors that influence situational awareness (CyberSA) for CyberOps (see Fig. 3, the second column marked in red border). These operational factors are derived using *attribute listing* and *morphological analysis* (MindTools, 2021). *Attribute listing and morphological analysis* allowed us to enumerate all operational attributes and features of each task, and by filtering against duplications, we removed factors that are either similar, duplicates or overlapped with existing features covered in other parts of the framework.

We observed that operational factors are influenced by technology factors when they are implemented correctly and ensure business factors are met appropriately. Operational factors range from tools and techniques employed to automate and orchestrated cybersecurity services to the policies and procedures that are leveraged to render these services efficiently and to the people that are tasked to operate and monitor the services, as shown in Fig. 3.

Note: This paper is focused only on the operational factors, we hope to cover the other factors and in addition to providing in-depth discussion on how the BOTH Factors are realised in future contributions (see Section 5 – *future research directions*).

The resultant operational factors are briefly discussed as follows:

### 4.1 Tools

These are the technical tools or technologies deployed in an organisation that allow operational aspects of the service to be swiftly executed and processed. Without capable tools in place, it will be challenging to realise operational efficiencies or meet some of the overarching business goals. For example, a security operations centre would be overwhelmed if they try to process or analyse high volumes of event logs without automation (e.g., parser, plugin, application programming interface) and technology (e.g., Security information and event management (SIEM)), because manual analysis of large corpse of logs will lag in time and consequently fail to achieve the business goal of real-time continuous monitoring. In addition, manual analysis is error-prone and less accurate, hence could impact the reliability and fidelity of the SOC analysis or outcomes.

Tools aid the SOC to become effectiveness in processing data and events (in realtime or near realtime), which helps CyberOps gain enhanced situational awareness of the current incident (situation), or impending and potential future manifestation of the same incident. It will also allow CyberOps to monitor and understand when there has been an escalation of the incident or when a similar incident occurs in future (a.k.a., predictive analytics).

Tools selection must be depending on achieving the features articulated by the Technology Factors (Onwubiko C., 2009) (as shown in Fig. 3), some of those include:
- Tools must be situation-aware,
- have the prerequisite interfaces,
- be automated allowing for orchestration and workflow processes.
- It should be integrated and support multiple interface types such as API, native etc., and should be smart and intelligent.
- It is important that operational tools have the capability to process huge amounts of data, easy to use, and portable. These are some of the features that guide tools selection and choice.

### 4.2 Techniques

This relates to the approaches CyberOps use to operationally identify, protect, detect, analyse, process and respond to cyber threats and cyber-attacks and remediate and recover from security breaches and exploits. CyberOps tasks and responsibilities are reliant on efficient techniques leveraging automation and machine intelligence for pace, speed, accuracy and precision.

For example, such techniques might be approaches to a Zero Trust architecture, endpoint protection, SOC operation, data science analysis, cyber event processing or incident management coordination. Irrespective of the chosen technique, we argue that techniques that allow CyberOps to leverage efficiencies in automation, workflow and orchestration offer the much-needed situational awareness of the organisation's DAAS, allowing operators to gain awareness, understanding and comprehension of immediate dangers, errors, events and incidents, and their future state change, which must also be understood. While speed, pace and accuracy are afforded by technology, automation and orchestration, but without situational awareness, better understanding and comprehension of the situation, then it is challenging for the operators to make informed and evidence-based risk decision over the various competing situations.

CyberOps techniques utilise industry best practices such as the MITRE ATT&CK framework (Mitre, 2017), CoCoa – an ontology-based and knowledge graph (Onwubiko C., 2018), the Lockheed Martin Cyber Attack Kill-chain (Lockheed Martin, 2016), CREST Cyber Threat Intelligence Maturity Assessment Tool (CREST, 2021), CHECK Tailored Security Testing (CHECK, 2021), Cyber Operational Recovery Framework (CRF, 2020) as guardrails for benchmarking and assessment.

Operational techniques must be driven by operational efficiency, performance, speed, accuracy and precision.

### 4.3 Architecture

Operational architectures are 'live' architectures in production and are used to operate, process and execute live services and systems. They include patterns, designs, interfaces and building blocks (i.e., architecture artifacts) that underpin business and operational technologies. Operational architectures for systems and services execute business requirements of the organisation to ensure the business needs are achieved. Operational architectures must use baselined architecture patterns, architecture designs and architecture blueprints that have been tested, approved and that are in use by the organisation.

We gain operational efficiencies and time savings by reusing existing architecture artifacts, and by ensuring continuous assurance of architecture contents. This means architecture repository is regularly updated with new and emerging secure by design architecture collaterals and artifacts. Adopting continuous integration, continuous development and continuous maintenance (CI, CD & CM) DevSecOps lifecycle, that is (Development, Security & Operation) methodology allows CyberOps to gain enhanced situational awareness of the architectures. It is important to note that infrastructures and architecture can now be instantiated in code, for example, infrastructure as code, or architecture as code, this is the case where infrastructure and architectures can be instantiated in code. In a pure cloud environment, infrastructures, such as EC2 (Elastic Cloud Compute) or virtual machines and virtual networking infrastructures can be instantiated in code, using cloud templates, e.g., AWS CloudFormation to do this.

Risks to architectures (for example, accidental human errors, misconfiguration and vulnerabilities in code or infrastructure) are minimised through continuous maintenance, using the DevSecOps pipeline. This is to ensure code and architecture are continually assured, and obsolete codes are removed, while new stacks are certified, reviewed and approved through automation and appropriate code review channels.

As discussed in Section 2 of this paper, by leveraging ZTA principles, through secure micro-segmentation, continuous trust and risk assessment of the identity (a.k.a., conditional access), and continuous entity and user behavioural analytics enhanced situational awareness of the monitored DAAS to be gained and maintained. Further, by collecting and understanding metrics around availability of service, continuity of service, reliability, performance of products and application, all of these can help enrich the overall CyberSA picture for operators.

Operational architectures must describe tasks that they accomplish, operational elements and information flows and patterns that should be used for (Dictionary of Military and Associated Terms, 2005), and they must be continuously reviewed and managed through architecture change boards. This is so that changes in the architecture space have full audit trail, authorisation and accountability.

### 4.4 Integration
To achieve foresight (for example, situational awareness) of any kind, cyber or otherwise, systems integration and automation are key. Operational technologies and systems ought to be integrated so that they form a cooperative and co-existing system of systems that deliver the overarching functionalities, interaction and business processing. With systems integration, disparate and diverse systems, components and subsystems that would have ordinarily existed as separate, isolated and siloed systems are interlaced as cooperative and coordinated services.

Cyber defences form a layer of protection (defense in depth) only when they are integrated. Integration can be achieved in hardware, software, programming and using hybrid methods, for example, systems integration of CPU to motherboard, keyboard and monitor is achieved through hardware integration, network integration can be hardware or software, process integration can be achieved through application programming interfaces (API), while robotic process automation can be achieved in hardware, software and hybrid (e.g., cyber physical systems).

Integration and automation of CyberOps is an absolute business requirement and considering the plethora of systems and applications that are used in CyberOps (see Fig. 1), it will be challenging, if not impossible, to monitor such myriad of systems, applications and networks without systems integration, process and technology automation. Systems integrations offer service efficiency and customer value-add through improved product quality and performance (Vonderembse M. A. et al, 1997). To gain enhanced situational awareness of CyberOps, integration and automation are extremely essential. Through these, information, interactions and awareness of current situations are learned, identified and comprehended, while impending and future situations can be predicted.

## 4.5 Automation

CyberOps relies on automation for speed, accuracy and precision. Without automation it will be challenging for CyberOps to be effective, relevant and responsive. For example, Cyber Ops depends on automation of tools, processes and procedures to analyse humongous volumes of logs and security event data at pace to gain situational awareness of current risks, impending situations and changes and evolution in future states of the incident (predictive analytics).

Automation allows cybersecurity operations to leverage machine intelligence, workflow automation, advance analytics and machine learning for information risk management, decision making, pattern matching, rule-based expert systems, conversational agent inference and extended detection and response (XDR) capability. Orchestration, tools automation, business process automation, workflow, and analysis offer fundamental enhancements to CyberOps for attack detection, analytics, incident management, investigation and forensics.

Advances in automation means CyberOps can become sophisticated in its responses, it can now stop active cyber-attacks (e.g., by sending) automated realtime response, for instance, TCP teardown, block malicious traffic, stop the execution of active malicious codes, automatic traffic re-routes and active cyber defence (ACD). These capabilities are challenging without automation, and it is inconceivable to operate CyberOps without automation, and where this is the case, then CyberOps will lag in time and would not be fit for purpose.

### 4.6 Coordination

Coordination is a human factor aspect that deals with the organisation of the different and disparate elements of a complex activity to enable them to work together coherently and effectively (Dictionary, 2019). Coordination is an intrinsic human cognitive function through which elements of complex tasks, operations and maneuver are arranged, organised, fused and/or managed collectively and collaboratively to achieve a common and desired goal.

Coordination is applied to both cyber and non-cyber related activities. Since this paper is focused on cyber situational awareness, our examples and explanations are drawn from the cyber domain, where coordination is an extremely valuable feature.

Take a SOC as an example, in the event of a cyber incident, security analysts are relied upon to analyse, investigate and coordinate cyber incident response, digital forensics, cyber-attack analysis and cyber incident management etc. These tasks on their own are complex, timely and cognitively demanding; allowing for actionable intelligence to be driven and appropriate mitigation to be executed to either stop the incident or minimise the impact in a coordinated fashion. Each aspect of cyber is interdependent, and it is this interdependence that necessitates coordination. For instance, responding to cyber incidents requires that activities of the SOC, such as monitoring, evidence gathering are coordinated with those of the cyber incident responders who perform activities such as gathering, preparation of digital evidence and the preservation of digital evidence, with the incident management helpdesk, and the senior management team who make decisions on the approach and possible cause of action (CoA), and including authorisation for reporting of the breach to national authorities, where applicable.

While operators may employ technology and automation to perform many of the complex tasks, allowing and leveraging machine intelligence, speed and accuracy, coordination is, and will remain largely a human factor attribute.

### 4.7-9 Policy, Process & Procedure

Policy, process and procedure are the foundation for CyberOps, without which it will be infeasible to conduct operational tasks. Policy provides guidelines, procedure stipulates the low level, 'how' of applying the policy, while the process helps industrialise the procedure and therefore allows a consistent approach to be followed. Cyber operation is no different. For example, they need policies, processes and procedures to operate the service, these may include policies on a wide range of tasks, from simple to complex tasks, such as joiners, movers and leavers (JML), access provision and deprovision, user rights management etc. The need procedures to follow, for example, cyber incident response procedure, major incident management procedure, etc., and likewise, they must have processes in place that allow consistency across various repetitive tasks, at least, for example, cyber incident playbook, access requisition process, account creation process etc.

The operational policies, processes and procedures provide a systematic and consistent guidance that allow for service efficient, improvement and performance. To stay ahead of the game, we argue that CyberOps must have very robust policies, procedures and processes and these need to be relevant, current and maintained, and most importantly, made readily available to staff. Often, process collaterals exist but staff are not aware of them due to poor communication or limited document management access or both. It is pertinent to note that the relevance of these collaterals depends on a number of factors, namely:

> a) All staff must be made aware that policies exist
> b) All staff should have access to policy collaterals
> c) All staff should be trained on the use of policy artefacts
> d) All staff must be briefed of the consequences abuse of policy.

### 4.10 People

People comprise staff (see Section 2, e.g., architects, engineers, analysts, administrators, scientists and management) who are responsible and accountable for engineering, administering, maintaining & supporting the operational day to day activities of the organisation ensuring that IT & OT systems and applications are effective, and perform in accordance to stipulated functional and business requirements.

There are many categories of operational staff, ranging from cyber operators, data scientists, cyber incident responders, administrators, analysts through to management teams. These people are incredibly important with the overall operational, administrative and change management aspects of the service operations and maintenance. They perform the business-as-usual tasks, operate the technologies that drive the processes and ensure that the tools and technologies are maintained, operated and serviced.

People are a critical aspect of CyberOps. While recent advances in technology, e.g., automation, machine intelligence and machine learning are useful and have helped improve CyberOps efficiencies and effectiveness, however, people are key. For example, if cyber defence systems are not continuously updated or patched, vulnerabilities may exist and this could result in the safeguards being exploited and further, they may then be used to compromise the wider ecosystem.

People aspect comes with its own challenge. Hiring highly skilled people is always a challenge and this has been exacerbated by the skills shortage in the cyber domain. In addition, the Tech expansion (increasing number of Startups and Enterprises) has increased competition for the sought-after employees resulting in staff retention also being a considerable challenge.

People (human operators) conduct cyber incident and crisis management, monitor the infrastructures, networks and systems, carryout analysis and investigations when a security breach occurs, and take part in decision making, escalations and reporting. People will always be required in most manner of endeavors in some form or another. This is most pertinent with cyber; while we now develop machine learning models that can predict cyber-attacks, artificial intelligence models that can recognise speeches and deep learning models that can investigate, recommend and optimise choices for humans, there are still, at least for now, areas and use cases where human operators are needed and will still be required in future. For example, in decision making, escalations, cyber security investigations, and prosecutions etc. There is no doubt that human operators will depend on cyber physical systems, machine learning models etc. for swifter, more precise and optimised choices, however, it will be a case of interdependency than replacement or displacement. We see a cooperative situation where humans leverage technological power, prowess, speed and accuracy in human decision making, prioritisation and resolution.

# 5    FUTURE RESEARCH DIRECTIONS

The contributions in this paper focused specifically on operational factors that influence CyberSA in cybersecurity operations. This is just one aspect of a main theme of work, we coined BOTH (Business, Operations, Technology and Human) factors. In future directions of this research, we hope to investigate the other factors. We plan to discuss Business, Technology & Human factors that influence CyberSA in CyberOps in future contributions.

# 6    CONCLUSIONS

CyberOps is an essential function of any organisation, be it a government department, industry, or academia. Regardless of the domain in which the cyber capabilities are applied to, for instance, aviation, maritime, intelligence community, industrial control systems (operation technology) or ICT, CyberOps must be governed, managed, operated and maintained.

CyberOps operators, namely engineers, administrators, analysts and scientists leverage operational factors to become efficient and effective on their functions. Operational factors help the operators not only to gain meaning situational awareness of the services they protect, but also, useful intelligence of the adversary which they must protect against. Operational factors for cyber situational awareness investigated encompass tools, techniques, integration, architecture, automation, coordination, policy, process and procedure and people.

To gain an enhanced cyber situational awareness it must be through a cooperative endeavor based on human to system (H$\rightarrow$S) relationships, as reliance on a singular aspect, say machine alone (without humans), will not provide the required levels of foresight; most pertinently, a relationship across multiple domains, which allows situational awareness of the ecosystem to be gained through monitoring, coordinating and responding to cyber incidents, leveraging system, process and task integration, automation, coordination and processing in a coherent and consistent manner that offer insight is required.

## BIOGRAPHICAL NOTES

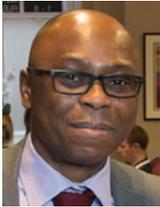 **Dr Cyril Onwubiko** is a Senior Member of the IEEE, and currently a Distinguished Speaker (DVP) of the IEEE Computer Society. Cyril is Director, Enterprise Security Architecture at Pearson, and also Director, Artificial Intelligence, Blockchain and Cyber Security at Research Series Limited, London, UK, where he directs Artificial Intelligence, Blockchain and Cyber Security programmes. Prior to Research Series, Cyril had worked in the Financial, Telecommunication, Health & Government and Public services Sectors. He is a leading scholar in Cyber Situational Awareness (Cyber SA), Cyber Security, Security Information and Event Management (SIEM) & Data Fusion, he has authored a couple of books including "Security Framework for Attack Detection in Computer Networks", and edited several books including "*Situational Awareness in Computer Network Defense: Principles, Methods & Applications*". Cyril is the Editor-in-Chief of the International Journal on Cyber Situational Awareness (IJCSA), and founder of the Centre for Multidisciplinary Research, Innovation and Collaboration (C-MRiC). He holds a PhD in Computer Network Security from Kingston University, London, UK; MSc in Internet Engineering, from University of East London, London, UK, and BSc, first class honours, in Computer Science & Mathematics.

## REFERENCE

**Reference to this paper should be made as follows**: Onwubiko C. (2020). CyberOps: Situational Awareness in Cybersecurity Operations. *International Journal on Cyber Situational Awareness*, Vol. 5, No. 1, pp82-107